\documentclass[12pt]{article}
\usepackage{graphicx}
\begin{document}
version 31.5.2015
\begin{center}
{\Large Explicit Bound State Description of Particles, \\Application to $p-e^-$ and $e^+-e^-$ Systems}   
\vspace{0.3cm}

H.P. Morsch \\ HOFF, Brockm\"ullerstr.~11,
D-52428 J\"ulich, Germany\\  E-mail:
h.p.morsch@gmx.de 
\end{center}

\begin{abstract}
Based on a QED Lagrangian with additional photon-photon coupling an explicit bound state description is presented, attempting a physical correct and parameter free description of free particles. Applied to $p-e^-$ and $e^+-e^-$ systems, with a "harmonic" boundary condition the deduced binding energies are consistent with Coulomb energies and radii in general agreement with other models. The sum of partial coupling strengths is in good agreement with $\alpha_{_{QED}}\sim$ 1/137, showing that this important constant can be deduced from first principles.

PACS/ keywords: 3.50.Kk, 11.15.-q, 31.15.Ne/ Explicit bound state
description of particles, based on a Lagrangian with Maxwell term, boson-boson and
boson-fermion coupling. Description of hydrogen and positronium bound states.  
Consistency of the integrated coupling strength with $\alpha_{_{QED}}$.
\end{abstract}

The study of fundamental forces is important to gain insight into the basic
structure of matter. For a satisfactory description of these forces a quantum field theory is needed, in which {\bf all} parameters can be derived from first
principles (completeness). This appears to be possible only for electromagnetic forces. For light atomic systems quantum electrodynamics (QED) gives rise to a quantitative description of spectra (by use of the Coulomb potential), fine and hyperfine structure splittings and Lamb shift as well as magnetic moments of leptons. Only one parameter is needed, the coupling constant $\alpha_{_{QED}}\sim$ 1/137, which is precisely determined from experimental data. However, a principal problem is that this parameter cannot be determined theoretically, because QED is an effective theory. The effective character is clearly visible in the structure of the Coulomb potential, a bound state potential of fermions.  However, a free bound state of nature (which has static as well as kinetic energy) cannot be composed of fermions only. Its kinetic energy gives rise to rotation, which would be spurious, if the fermion recoil could not be absorbed by other particles (photons). This requires that photons are not only the source of the interaction (boson-exchange) but have to stabilize the dynamics of the fermions as well. Thus, a free particle must have a double bound state structure of fermions {\bf and} bosons. 
This property has to be included in an explicit and physically correct bound state version of QED, from which one can hope to deduce also the parameter $\alpha_{_{QED}}$ from basic considerations. Only then QED can be considered as a fundamental {\bf and} complete theory. 

Because of these arguments an explicit bound state version of QED must have a Lagrangian of more complex structure than the usual first order QED Lagrangian~\cite{gaugeth} with additional boson fields, which balance the motion of fermions. The evaluation of such a Lagrangian is more tedious, but it leads to a finite theory (with all advantages over effective and divergent theories) and should lead to a real physical understanding of the mechanisms involved. This paper describes an application of this formalism to the atomic systems $p-e^-$ and $e^+-e^-$. 

The Lagrangian with fermions of masses $m_1$ and $m_2$ is of the form 
\begin{equation}
\label{eq:Lagra}
{\cal L}=\frac{1}{\tilde m^{2}} \bar \Psi\ i\gamma_{\mu}D^{\mu}
D_{\nu}D^{\nu}\Psi\ -\ \frac{1}{4} F_{\mu\nu}F^{\mu\nu}~,   
\end{equation}
where $\tilde m$ is the mass parameter $\tilde m=m_1 m_2/(m_1+m_2)$ and
$\Psi$ are charged fermion fields, $\Psi=\Psi^+$ and $\bar \Psi= \Psi^-$.   
Vector boson fields $A_\mu$ with coupling $g$ to fermions are contained in the
covariant derivatives $D_{\mu}=\partial_{\mu}-i{g} A_{\mu}$. 
The second term of the Lagrangian represents the Maxwell term with
Abelian field strength tensors $F^{\mu\nu}$ given by $F^{\mu\nu}= 
\partial^{\mu}A^{\nu}-\partial^{\nu}A^{\mu}$, which gives rise to both electric 
and magnetic coupling. 

This Lagrangian includes naturally higher order boson and fermion fields. In the past two arguments have been brought forward against the use of this type of Lagrangian: the necessary $1/\tilde m^2$ factor should give rise to uncontrolled divergences in standard (infinite) gauge theories; further, a Lagrangian with higher order fermion fields will lead to nonphysical solutions~\cite{highd}. However, both arguments are not valid in the present case (in which the inclusion of boson-boson coupling is absolutely necessary): the Lagrangian leads to a finite theory; further, in the present formalism non-physical solutions can be excluded by strict geometrical and energy-momentum constraints. 

By inserting $D^{\mu}=\partial^{\mu}-i{g} A^{\mu}$ and
$D_{\nu}D^{\nu}=\partial_{\nu}\partial^{\nu}
-ig(A_{\nu}\partial^\nu+\partial_\nu A^\nu) -g^2 A_\nu A^\nu$ in
eq.~(\ref{eq:Lagra}), the first part of ${\cal L}$ gives rise to a number of
terms, which contain boson and fermion fields and/or their
derivatives. All terms containing the derivative of the fermion fields $\partial^{\nu} \Psi$ are related to a complex dynamics of the system. For stationary solutions only two terms of the Lagrangian contribute
\begin{equation}
\label{eq:L2}
{\cal L}_{2g} =\frac{-ig^{2}}{\tilde m^{2}} \ \bar \Psi~\gamma_{\mu}
[A^\mu \partial_{\nu} A^\nu]~\Psi \ 
\end{equation}
and 
\begin{equation}
\label{eq:L3}
{\cal L}_{3g} =\frac{ -g^{3}\ }{\tilde m^{2}} \ \bar \Psi~\gamma_{\mu}
[A^\mu A_\nu A^\nu]~\Psi \ .
\end{equation}
As gauge condition we use $\partial^2 A^\nu=0$. 

From the Lagrangians~(\ref{eq:L2}) and (\ref{eq:L3}) fermion matrix
elements have been derived, a standard method based on generalized Feynman diagrams, see e.g.~ref.~\cite{gaugeth}. These have been used in 
the form ${\cal M}^f=<g.s.|~K(p'-p)~|g.s.>\sim \bar \psi(p')\ K(q)~\psi( p)$, where $\psi(p)$ is a fermion wave function $\psi(p)=\frac{1}{\tilde m^{3/2}} \Psi(p_1)\Psi(p_2)$ and $K(q)$ a kernel related to the boson structure of the Lagrangian. In the present case it is given by 
$K(q)=\frac{1}{\tilde m^{5}}\ [O^3(q_i)\ O^3(q_j)]$, in which $O^3(q_i)$ represents
a product of boson fields or derivatives given by the square brackets in eqs.~(\ref{eq:L2}) and (\ref{eq:L3}). Using $\alpha=g^2/4\pi$ this leads to matrix elements of the form 
\begin{equation}
\label{eq:M2}
{\cal M}_{2g} =\frac{\alpha^{2}}{\tilde m^5}~\bar \psi(p')~\gamma_\mu
A^{\mu}(q_2)~(\partial_\nu A^{\nu}(q'_2))(\partial_\sigma
A^{\sigma}(q'_1))~\gamma_\rho A^{\rho}(q_1)~\psi(p)\ 
\end{equation}
and
\begin{equation}
\label{eq:M3}
{\cal M}_{3g} = \frac{-{\alpha}^{3}}{\tilde m^5}~\bar
\psi(p')~\gamma_\mu A^\mu(q_2)~A_\nu(q'_4)A^\nu(q'_3)
A_\sigma(q'_2)A^\sigma(q'_1)~\gamma_\rho A^\rho(q_1)~\psi(p) \ . 
\end{equation} 
One may compare these matrix elements to similar ones derived from the first order QED Lagrangian ${\cal L}_{f.o.}= \bar \Psi\ i\gamma_{\mu}D^{\mu}
\Psi\ -\ \frac{1}{4} F_{\mu\nu}F^{\mu\nu}$. By writing similarly ${\cal M} = \bar \psi(p')\ K(q)~ \psi(p)$, with $K(q)=\frac{1}{\tilde m}\ [O^1(q_2)\ O^1(q_1)]$ one obtains for the case  $\partial\Psi=0$ only one (boson-exchange) matrix element ${\cal M}_{f.o.} = \frac{-{\alpha}}{\tilde m}~\bar \psi(p')~\gamma_\mu A^\mu(q_2)\gamma_\rho A^\rho(q_1)~\psi(p)$. Since the boson-fields $A^\mu(q_i)$ are relativistic, they overlap only momentarily and cannot form a stable potential. Only in the non-relativistic limit (which is not realized for strongly bound atomic states) one could write ${\cal M}=\bar \psi(p')~V(q)~\psi(p)$, where $V(q)\sim\alpha\cdot 1/q^2$ is the Coulomb potential. 

The comparison of both theories shows {\bf two essential} differences, important for a correct physical description of particle bound states: 1. The "boson-exchange" matrix element ${\cal M}_{3g}$ has a more complex structure than ${\cal M}_{f.o.}$ with additional boson fields, needed to balance the fermion motion. 2. A second matrix element ${\cal M}_{2g}$ is present, which does not exist in first order theories. This term leads to a dynamical stabilization (confinement) of the system (discussed below). 

From these matrix elements bound state potentials can be deduced. 
First we replace in eq.~(\ref{eq:M2}) the bosonic part
$(\partial_\nu A^{\nu}(q'_2)) (\partial_\sigma A^{\sigma}(q'_1))$ for $\nu =\sigma$ by $\frac{1}{2}\partial^2 (A_\nu(q'_2) A^\sigma(q'_1))$. For $\nu \neq \sigma$ there are strong cancellations and the corresponding matrix element has been neglected. Then (analogue to the fermion wave functions) normalized boson (quasi) wave
functions of scalar ($\mu =\nu$) and vector ($\mu \neq\nu$) structure are
introduced $W_{\mu}^\nu(q')= \frac{1}{\tilde m} A_\mu(q'_j)A^\nu(q'_i)$. Further, a boson-exchange interaction is obtained with a form $V_{\mu}^\nu(q)= \frac{1}{\tilde m} A_\mu(q_2)A^\nu(q_1$) ($\mu \neq\nu$), which is similar to first order QED. 
The fact that boson fields can be combined to normalized wave functions, leads quite naturally to a finite theory. 

By equal time requirement the fermion and boson vectors can be 
reduced by one dimension, yielding boson wave
functions\footnote{with dimension $[GeV]$.} of scalar and vector structure
$w_{s}(q')$ and $w_{v}(q')$ and an interaction  $v_{v}(q)$.
This yields 
\begin{equation}
\label{eq:M2s}
{\cal M}_{2g} =\frac{\alpha^{2}}{2\tilde m^{3}}~\bar \psi(p')
\ w_{s}(q')\ \partial^2 w_{s}(q')\ \psi(p)\ 
\end{equation}
and 
\begin{equation}
\label{eq:M3s}
{\cal M}_{3g} = \frac{-{\alpha}^{3}}{\tilde m^2}~\bar
\psi(p')~w_{s,v}(q')v_{v}(q)w_{s,v}(q')~\psi(p) \ .
\end{equation}
The bosonic part of eq.~(\ref{eq:M3s}) can also be written in the form
of a matrix element, in which the wave functions $w(q')$ are
connected by $v_v(q)$
\begin{equation}
\label{eq:P2g'}
{\cal M}^{g} =\frac{-\alpha^{3}}{\tilde m^{2}}
\ w_{s,v}(q')~v_{v}(q)~w_{s,v}(q') .
\end{equation}

In the following an attempt is made to evaluate these matrix elements. We rely on the Hamiltonian formalism by relating kinetic and potential energies by $(T+V)\psi=E \psi$. Further, binding energies have been evaluated by using the virial theorem. Finally energy-momentum conservation is assumed, which is known to be valid for relativistic systems. If these conditions would not be realized, reasonable results could not be expected.

Going to r-space the fermion matrix element~(\ref{eq:M2s}) can be written by 
\begin{equation}
\label{eq:P2g}
{\cal M}_{2g} = \bar \psi(r)\ V_{2g}(r)\ \psi(r)\ ,
\end{equation}
in which $V_{2g}(r)$ is a potential, which can be derived from a boson Hamiltonian of a form
\begin{equation}
\label{eq:H}
-\frac{\alpha^2 (\hbar c)^2}{4\tilde m}~\Big (\frac{d^2
    w_s(r)}{dr^2} + \frac{2}{r}\frac{d w_s(r)}{dr}\Big ) +V_{2g}(r)~w_s(r)
  = E_i~w_s(r)~.
\end{equation}
This leads to
\begin{equation}
V_{2g}(r)= \frac{\alpha^2 (\hbar c)^2}{4\tilde m}\ \Big
(\frac{d^2 w_s(r)}{dr^2} + 
  \frac{2}{r}\frac{d w_s(r)}{dr}\Big )\frac{1}{\ w_s(r)}+E_o\ .
\label{eq:vb}
\end{equation}
A connection to the vacuum is made by assuming $E_o=E_{vac} = 0$. This potential is of large importance, since it leads to dynamical stabilization and confinement of the system: with positive eigenvalues fermion-antifermion pairs are locked in this potential during overlap of boson fields and form a stable system, which cannot decay.  $V_{2g}(r)$ shows a quite linear rise towards larger radii, very similar to the empirically introduced confinement potential in hadron potential models~\cite{qq}. 

Further, the matrix element~(\ref{eq:M3s}) can be written in r-space by
\begin{equation}
\label{eq:M3f}
{\cal M}_{3g} = \bar \psi(r)\ V_{3g}(r)\ \psi(r) \ ,
\end{equation}
in which the potential $V_{3g}(r)$ has the form of a folding potential 
\begin{equation} 
\label{eq:vqq}
V_{3g}(r)= -\frac{\alpha^3 \hbar c}{\tilde m} \int dr'\ 
w_{s,v}(r')\ v_v(r-r')\ w_{s,v}(r')~   
\end{equation}
with an interaction $v_v(r)=-\hbar c~w_v(r)$. As mentioned above, this potential can also be considered as boson matrix element, in which the bosons are "bound" in the potential $v_v(r)$.

The structure of ${\cal M}_{3g}$ gives rise to two states (scalar and vector) without angular momentum (L=0) and boson wave functions $w_{s,v}(r)$. The corresponding fermion wave functions $\psi_{s,v}(r)$ have to be of similar radial form $\psi_{s,v}(r) \sim w_{s,v}(r)$. These are orthogonal, leading to the constraint 
\begin{equation}
\label{eq:ortho}
\int r^2dr~\psi_s(r) \psi_v(r)=\int r^2dr~w_s(r) w_v(r)=<r_{w_s,w_v}>=0 \ .
\end{equation}
To satisfy this condition, for a given wave function of the scalar state $w_{s}(r)$ that of the vector state can be written in the form 
\begin{equation}
\label{eq:spur}
w_{v}(r) = w_{v_o}~[w_s(r)+\beta R\ \frac{d w_s(r)}{dr}]~, 
\end{equation}
where $w_{v_o}$ is obtained from the normalisation $2\pi \int r dr\ w_v^2(r)
=1$ and $\beta R$ is given by $\beta R=-\int r^2dr~w_s(r)/\int r^2dr~[d
  w_s(r)/ {dr}]$. Because of the derivative structure $w_{v}(r)$ has a smaller root mean square radius than $w_{s}(r)$. Therefore, a natural geometric 
condition requires that the interaction for this state takes
place inside the bound state volume of $w_s^2(r)$. This leads to the geometrical boundary condition
\begin{equation}
\label{eq:conr}
|V^v_{3g}(r)| \simeq {c}\ w^2_s(r)  \ . 
\end{equation}

The conditions~(\ref{eq:ortho}) and (\ref{eq:conr}) require a form of  
boson wave function of the scalar state 
\begin{equation}
\label{eq:wf}
w_s(r) = w_{s_o}\ exp\{-(r/b)^{3/2}\} \ , 
\end{equation} 
where $w_{s_o}$ is fixed by the normalisation $2\pi \int r dr\ w_s^2(r)
=1$. The slope parameter $b$ as well as the coupling constant
$\alpha$ has to be determined from boundary conditions as discussed below. 

In addition to states with L=0 also two states with
angular momentum L=1 (p-states) exist, for which similar forms of their wave functions can be assumed. In atomic systems all L=0 and L=1 states give rise to degenerate singlet and triplet states. However, fine and hyperfine structure splittings of these states are observed, which are in the hydrogen atom 5-6 orders of magnitude smaller than the binding energies. These splittings, as well as very small shifts (as the Lamb shift), are satisfactorily described in QED and are not considered in the present analysis.

The general structure of the bound state solutions is shown in fig.~1 for a system with root mean square radius $<r_{w_s}^2>^{1/2}$ = 86 pm. In the
upper part the radial dependence of the interaction $v_v(r)$ is compared to
the $1/r$ dependence of the Coulomb potential, which shows that there are no divergences for $r\to$ 0 and $\infty$ in the present description. In
the middle part the radial 
dependence of boson density $w^2_s(r)$ and potentials $V_{3g}^{s,v}(r)$ is
shown, which indicates that relation (\ref{eq:conr}) is reasonably well
fulfilled. Only for large radii $w^2_s(r)$ falls off less rapidly than
$V_{3g}^{v}(r)$, which shows that a small mixing between $w_s(r)$ and $w_v(r)$
(in the order of 10-15 \%) is needed to satisfy eq.~(\ref{eq:conr}) at large radii. 
In the lower part the potential $V_{2g}(r)$ is displayed,
which shows a quite linear increase at larger radii expected for the confinement potential.  

Binding energies have been calculated by using the virial theorem in the radial form $E^{ng}_f= 4\pi [\int r^2dr~\psi^2(r) V_{ng}(r)
  -\frac{1}{2}\int r^3dr~\psi^2(r)\frac{d}{dr}V_{ng}(r)]$, where the fermion wave functions $\psi(r)$ are normalized by 
$4\pi \int r^2 dr\ \psi^2(r) =1$. In addition, $V_{3g}(r)$ can be interpreted 
as "bound state" of bosons. The corresponding binding energies $E_g$ have been calculated by $E_g=2\pi[\int rdr~w^2(r)v_v(r) -\frac{1}{2}\int
  r^2dr~w^2(r)\frac{d}{dr}v_v(r)]$.
   
Energy-momentum conservation requires that the (negative) 
binding energies of fermions and bosons $E^{3g}_{f}$ and $E_g$ are compensated
by the sum of their root mean square momenta  
\begin{equation}
<q^2_{f}>^{1/2}+<q^2_{g}>^{1/2} = -(E^{3g}_{f}+E_g)/c\   
\label{eq:massq}
\end{equation}
with boson momentum square $<q^2_{g}>=\int q^3dq~V_{3g}(q)/\int qdq~V_{3g}(q)$ and a similar quantity for fermions $<q^2_{f}>=(\int q^4dq~\psi^2(q)/\int q^2dq~\psi^2(q))<q^2_{g}>$.
This constraint stems from the requirement that equal properties should be found in r- and q-space. Condition~(\ref{eq:massq}) has to be fulfilled for both scalar and vector states. This may be taken as strict consistency check of the assumed wave functions.  

--------------

An application of this formalism is discussed for the atomic bound state systems $p~-e^-$ and $e^+-e^-$, which have been studied previously in the Bohr model, with the Schr\"odinger equation, the Dirac equation and in QED (using effective potentials). However, a fully relativistic gauge theory leading to a realistic bound state description has not been found.

First, s-states (without angular momentum, L=0) are discussed. The slope parameter $b$ can be determined by satisfying eq.~(\ref{eq:massq}). By increasing the slope parameter $b$ (and consequently also the root mean square radius $R_w=<r^2_w>^{1/2}$) the total momentum $q_t=<q^2_{V_{3g}}>^{1/2}+<q^2_{v_v}>^{1/2}$ decreases, as shown by the dot-dashed line in the upper part of fig.~2. Differently, the total binding energy $E_t=E^{3g}_{f}+E_g$ increases with $R_w$ (solid line), if $E_f$ is adjusted to the experimental binding energies. The constraint (\ref{eq:massq}) is fulfilled for the value of $R_w$ at which the two lines overlap.

Since $E_t$ depends on $b$ but also on the coupling constant $\alpha$, a careful analysis is needed to avoid ambiguities. For the hydrogen 1s and 2s states with binding energies of -13.6 and -3.4 eV a solution has been found with $b$=105 pm, $\alpha$=1.93 and a mixing of $E^s_f$ and $E^v_f$ of 10 \%. With $E_t$= -4.2 keV and $q_t$= 4.2 keV/c for the 2s state and $E_t$= -6.3 keV and $q_t$= 6.3 keV/c for the 1s state, energy-momentum conservation is satisfied for both states. 

Binding energies $E_f^{3g}$ of -17.9 eV and -8.9 eV are obtained for the 1s and 2s state, respectively, whereas the corresponding values of $E_f^{2g}$ are 4.1 eV and 13.6 eV. One can see that for the energy $E_{f_s}^{2g}$ of the scalar 2s state a reduction by a factor of about 2 is needed to get agreement with the experimental binding energies (with a mixing of the wave functions of both states of about 10 \%). A possible explanation of this reduction (only for scalar states) is that the derivative structure of $V_{2g}(r)$ couples much weaker to scalar than to vector states,  in the ratio 1/3 according to a (2s+1) factor. The justification for such a refinement is still needed.  

The radial properties of the resulting density and potentials are shown in fig.~1.
With a matching of energy and momentum as shown in fig.~2, the root mean square radii $<r_{w_{s,v}}^2>^{1/2}$ are found to be 86 and 49 pm for scalar and vector state, respectively, with estimated uncertainties of 10-15 \%. Since $w_s(r)$ match the radial form of the vector potential $V^v_{3g}(r)$ by the  condition~(\ref{eq:conr}), the root mean square radius of the 1s potential $<r_{V_{1s}}^2>^{1/2}$ is 86 pm, leading to a radius at half maximum of $R^{1s}_{1/2}$ of 53 pm, in good agreement with the radii deduced from other models, see table~1. 

It is interesting to see, in which way energy-momentum conservation is fulfilled in the present system of fermions and bosons. For the 2s state the average boson momentum $<q^2_g>^{1/2}$ of 4.2 keV/c is about 3 orders of magnitude larger than the average fermion momentum. Similarly, the binding energies show a strong imbalance between fermions and bosons with $E_f$= -3.4 eV and $E_g$= -4.2 keV. This indicates that energy-momentum conservation is entirely realized by bosons.   

\begin{table} [t]
\caption{Results for $p - e^-$ and $e^+-e^-$ bound state solutions n, using $\alpha$=1.93 and factor 2 reduction of $E_{f_s}^{2g}$. Binding energies $E_f$ are given in eV, $b$ and radii in pm.}  
\begin{center}
 $p-e^-$ \\
\begin{tabular}{c||ccc|cc|cc}
sol. & $E_f$(ns) & $E_f$(2ns, 2np) & $E_f$(4np) & $b$ & $R^{~ns}_{1/2}$ & n $R_{Bohr}$ & $R_{cov}^{~*}$ \\  
\hline
~1  & -13.6 (1s) & -3.4 ~(2s) (2p) & -0.85 (4p) & 105 & ~53 & ~53 & 31$\pm$5\\ 
~2  & -3.4 ~(2s) & -0.85 (4s) (4p) & -0.21 (8p) & 210 & 105  & 106 \\ 
~3  & -1.51 (3s) & -0.38 (6s) (6p) & -0.09 (12p) & 315 & 158 & 159 \\ 
~4  & -0.85 (4s) & -0.21 (8s) (8p) & -0.05 (16p) & 420 & 211 & 212 \\ 
~5  & -0.54 (5s) & -0.14 (10s) (10p) & -0.03 (20p) & 525 & 264  & 265 \\ 
~6  & -0.38 (6s) & -0.09 (12s) (12p) & -0.03 (24p) & 630 & 316  & 318 \\ 
~7  & -0.28 (7s) & -0.07 (14s) (14p) & -0.03 (28p) & 735 & 369  & 371 \\ 
~8  & -0.21 (8s) & -0.05 (16s) (16p) & -0.03 (32p) & 840 & 422  & 424 \\ 
\end{tabular}

 $e^+-e^-$ \\
\begin{tabular}{c||ccc|cc|cc}
sol. & $E_f$(1$^-$) & $E_f$(1$^-$, 0$^+$) & $E_f$(0$^+$) & $b$ & $R_{1/2}^{~ns}$~ & n $R_{Bohr}$ & $R_{cov}^{~*}$ \\  
\hline
~1  & ~-6.8 (1s) & ~~-1.7 ~~(2s) (2p) & -0.43 (4p)~~ & 210~ & 105 & 106 & ~~~~~~~~ \\ 
~2  & ~-1.7 (2s) & ~-0.43~ (4s) (4p) & -0.11 (8p)~ & 420 & 211 & 212 \\ 
\end{tabular}
 \end{center}
$^*$ covariant radius from ref.~\cite{cov}.
\end{table}
%\vspace{0.5cm}
Other solutions exist for larger values of the slope parameter $b$. The next solution, shown in fig.~2, needs a value of $b$ of 210 pm (with $\alpha$ and factor 2 reduction of $E_{f_s}^{2g}$ unchanged), which is exactly the double of $b$ deduced for the first solution ($b_2=2b_1$). This gives rise to 2s and 4s states, again in good agreement with experiment. Other solutions are found for $b_n=nb_1$, where n are integers 3, 4, 5, ... Results for solutions up to n=8 (with energy-momentum matching given by dashed lines in fig.~2) are given in table~1. The deviations from the corresponding Coulomb energies are less than 1 \%. 

Solutions for p-states (with angular momentum, L=1) can be obtained with similar wave functions as for s-states. For solution 1 in table~1 a dominant wave of vector structure yields agreement with the binding energy of the 4p-state at -0.85 eV, whereas the 2p-state binding energy of -3.4 eV is obtained by a wave function dominated by scalar form. In a similar way also the other p-wave solutions in table~1 are obtained.
\vspace{0.3cm}

By applying the above formalism to the $e^+-e^-$ system, the mass parameter $\tilde m$ is a factor of two smaller than for the hydrogen atom. By keeping the radii unchanged, binding energies would be obtained, which are a factor of two larger than in $p-e^-$. To obtain agreement with the known positronium spectrum the radii have to be increased by a factor 2 (again $\alpha$ and factor 2 reduction of $E_{f_s}^{2g}$ unchanged). The resulting dependencies of the densities and potentials are very similar to those of the $p-e^-$ system in fig.~1, energy-momentum matching for this system is shown in the lower part of fig.~2 and results for the two lowest solutions are given in table~1.

The necessity of different radii for hydrogen and positronium, but also the existence of solutions with different binding energies indicates that for a self-consistent and complete description further boundary conditions are needed. One condition arises from the structure of the confinement potential~(\ref{eq:vb}), which (from dimensional arguments) can be written in a different form 
\begin{equation}
V_{2g}(r)= \frac{\alpha^2~\xi (E_f/2) <r^2_{w_s}>}{4 }\ \Big
(\frac{d^2 w_s(r)}{dr^2} + 
  \frac{2}{r}\frac{d w_s(r)}{dr}\Big )\frac{1}{\ w_s(r)} \ ,
\label{eq:vb1}
\end{equation}
where $\xi$ is an adjustment parameter.
This leads to   
\begin{equation}
Rat_{conf}=\frac{\xi \tilde m (E_f/2)<r^2_{w_s}>}{(\hbar c)^2} =1\ .
\label{eq:ratc}
\end{equation}
With $\xi$=6 this constraint is satisfied for all solutions in table~1; in particular, it requires for the $e^+-e^-$ system a radius of a factor 2 larger than for the hydrogen atom. 

Another constraint, very special for the systems in question, requires that all solutions with n=1, 2, 3, 4, ...~satisfy the "harmonic" condition for the slope parameters $b_n$
\begin{equation}
b_n= n~b_1 \ . 
\label{eq:bnn}
\end{equation}
The sum of partial strengths and energies (related to $1/b_n$) follow the harmonic series $1+1/2+1/3+1/4+..$. 
The development of higher harmonic modes may be caused by the fact that the energy ratios between scalar and vector states of a factor 4 match exactly the energies of the higher harmonics (see table~1). Such features cannot be expected for systems bound by other forces.
%\vspace{0.3cm}

With these constraints binding energies have been extracted, which are in perfect agreement with the Coulomb energies up to large n (with deviations by not more than 0.5-1 \% tested up to n=100), see fig.~3. Because of this close correspondence it should be possible to understand the magnitude of $\alpha_{_{QED}}$ from the present approach. To achieve this, partial coupling strengths have been calculated for each solution n, given by  $\alpha^n_{\Delta}=2\alpha^3~(\int dr V^{s,n}_{3g}(r))/(\int dr~\hbar/r)$, where the factor 2 is due to s and p contributions. The sum over all n should then be comparable to $\alpha_{_{QED}}$. 
In a first step the potentials $V^{n}_{3g}(r)$ have been calculated for each n independent of all other solutions n'$\neq$n. This leads to $\alpha^n_{\Delta}= \alpha^1_{\Delta}/n$, which is proportional to the harmonic series (see above) and yields $\sum_{n=1}^{~\infty}\alpha^n_{\Delta}$ divergent. The corresponding spectrum is identical to the Coulomb energy spectrum in fig.~3.
 
However, for higher harmonic modes an independence from solution 1 cannot be expected. Their potentials should follow the radial dependence of the potential $V^{s}_{3g}(r)$ for n=1. Using a weight function $\Omega_n=V^{s}_{3g}(r_n)/V^{s}_{3g}(r_1)$ with $r_n=1/n$, the potentials $\Omega_nV^{n}(r)$ result in binding energies for ns states shown in fig.~3 by open squares, which fall off more rapidly than the Coulomb energies (given by solid points). However, up to n=10 very small differences between $E_f(ns)$ and $E_{Coul}(ns)$ are found, which are difficult to detect experimentally.

Also the partial coupling strengths $\Omega_n \alpha^n_{\Delta}$ fall off significantly stronger than $\alpha^1_{\Delta}/n$ for large n (for n=400 the value of $\Omega_n$ is already fallen off by a factor 7 $10^{-6}$), leading to rapid convergence of $\sum_{n=1}^{~\infty}\Omega_n \alpha^n_{\Delta}$. This yields $\sum_{n=1}^{500}\Omega_n \alpha^n_{\Delta}=7.5~10^{-3}$, which is in excellent agreement with $\alpha_{_{QED}}\sim 7.3~10^{-3}$ within very small uncertainties. 

\vspace{0.5cm}

The present results may be summarized as follows: \\
1. The general arguments against the use of higher order Lagrangians (leading to divergences and ghosts) are not valid for the present Lagrangian. \\
2. A correct description of the dynamics of free particle bound states is obtained without spurious fermionic motion: the fermion recoil is absorbed by bound and therefore massive bosons. \\
3. The confinement potential $V_{2g}(r)$ warrants dynamical stability of the system. \\
These features are imperatively required for any free particle bound state. In particular, the electron must have such a fermion-photon structure. Therefore, as in first order QED Compton scattering has to scale with $1/p$, where p is the electron momentum. \\
4. A quantitative description of the $p-e^-$ and $e^+-e^-$ systems is obtained without open parameters. The validity of the Coulomb energy spectrum (up to $n\sim 10$) is confirmed. \\
5. The electric coupling (fine structure) constant $\alpha_{_{QED}}\sim 1/137$ is reproduced, supporting firmly the validity of the present approach. 

Apart from the first point all others speak {\bf against} the use of first order Lagrangians for a real understanding of free particle bound states: \\
1. A first order Lagrangian leads to spurious motion of fermions. \\
2. In relativistic cases stable bound state potentials do not exist. \\ 
3. The validity of the Coulomb potential for strongly bound states is not understood. \\
4. The electric coupling constant cannot be derived from first principles.

\vspace{0.5cm}
%\newpage

For fruitful discussions, direct help in the derivation of the formalism and
general support the author is indebted to many colleagues, in particular to
B. Loiseau, P. Decowski and P. Zupranski.

\begin{figure} [t]
\centering
\includegraphics [height=18cm,angle=0] {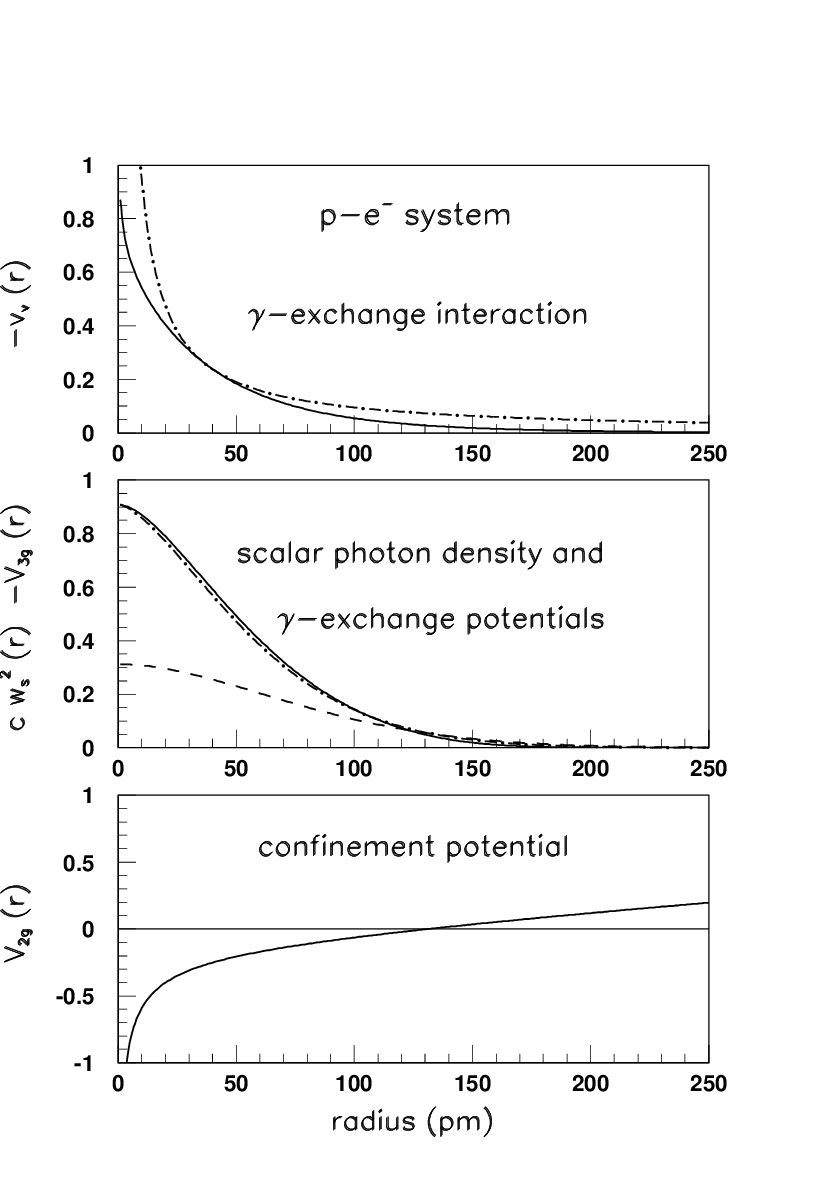}
\caption{Radial dependence  of a self-consistent solution for a $p - e^-$ bound state with $<r^2_{w_s}>^{1/2}$ = 86 pm. 
  \underline{Upper part:} Relative interaction $v_v(r)$ in comparison with the
  Coulomb potential given by dot-dashed line.
 \underline{Middle part:} Boson density $w_s^2(r)$ (dot-dashed line) and
 boson-exchange potentials $|V^{s,v}_{3g}(r)|$ given by dashed and solid
 lines, respectively. 
 \underline{Lower part:} Confinement potential $V_{2g}(r)$. }    
\label{fig:g1exep}
\end{figure} 

\begin{figure} [t]
\centering
\includegraphics [height=17cm,angle=0] {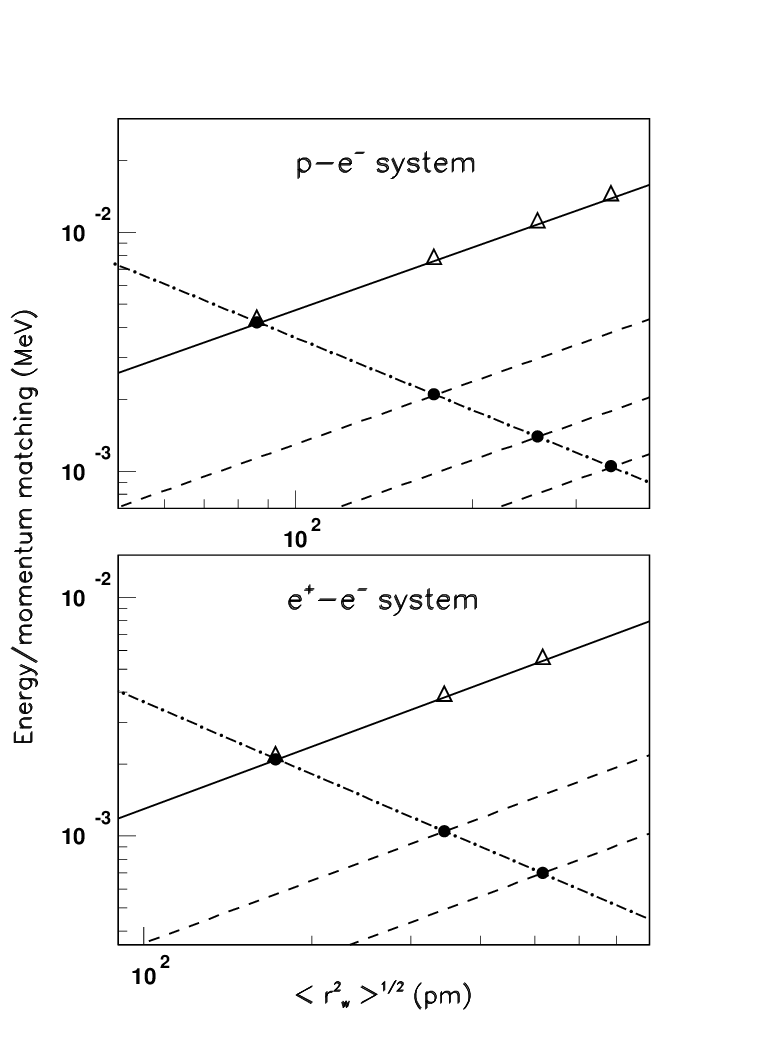}
\caption{Total momentum $q_t$ (solid points) and total binding energy $E_t$ (open triangles) for $p-e^-$ (upper part) and $e^+-e^-$ systems (lower part) as a function of $<r^2_{w_s}>^{1/2}$. 
Linear interpolations for $q_t$ are given by dot-dashed lines, those for $E_t$ for solution 1 by solid and for n$>$1 by dashed lines. }  
\label{fig:match}
\end{figure} 

\begin{figure} [t]
\centering
\includegraphics [height=17cm,angle=0] {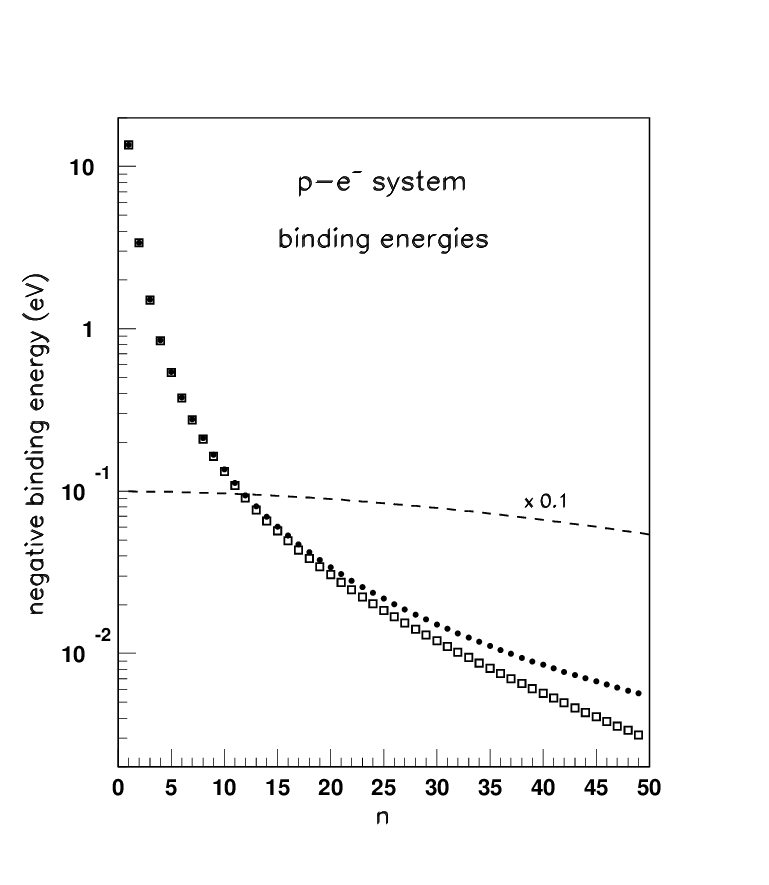}
\caption{Absolute values of the binding energies for ns states in the $p-e^-$ system as a function of n. The small closed points relate to the Coulomb energies, the open squares to $E_f^n$, yielding a sum of partial couplings strengths in agreement with $\alpha_{_{QED}}$. The correction factors $\Omega_n$ to the Coulomb energies are given by dashed line, which follow the radial dependence of the potential $V^s_{3g}(r)$.}    
\label{fig:epharm}
\end{figure} 

\end{document}